\documentclass[aps,prb,twocolumn,showpacs,superscriptaddress]{revtex4}
\usepackage{graphicx}
\usepackage{color}

\begin{document}
\title{Phase diagram of the Bose-Hubbard model on a ring-shaped lattice with tunable weak links}

\author{Kalani Hettiarachchilage} 
\affiliation{Department of Physics and Astronomy, Louisiana State University, Baton Rouge, Louisiana 70803, USA}
\author{Val\'ery G.~Rousseau}
\affiliation{Department of Physics and Astronomy, Louisiana State University, Baton Rouge, Louisiana 70803, USA}
\author{Ka-Ming Tam}
\affiliation{Department of Physics and Astronomy, Louisiana State University, Baton Rouge, Louisiana 70803, USA}
\affiliation{Center for Computation and Technology, Louisiana State University, Baton Rouge, LA 70803, USA}
\author{Mark Jarrell}
\affiliation{Department of Physics and Astronomy, Louisiana State University, Baton Rouge, Louisiana 70803, USA}
\affiliation{Center for Computation and Technology, Louisiana State University, Baton Rouge, LA 70803, USA}
\author{Juana Moreno}
\affiliation{Department of Physics and Astronomy, Louisiana State University, Baton Rouge, Louisiana 70803, USA}
\affiliation{Center for Computation and Technology, Louisiana State University, Baton Rouge, LA 70803, USA}

\date{\today}

\begin{abstract}
Motivated by recent experiments on toroidal Bose-Einstein condensates in all-optical traps with tunable weak links, 
we study the one-dimensional Bose-Hubbard model on a ring-shaped lattice with a small region of weak hopping integrals using 
quantum Monte Carlo simulations.  Besides the usual Mott insulating and superfluid phases, 
we find a phase which is compressible but non superfluid with a local Mott region. This `local Mott' phase
extends in a large region of the phase diagram. These results suggest that the insulating and conducting phases 
can be tuned by a local parameter which may provide a new insight to the design of atomtronic devices.
\end{abstract}

\maketitle

\paragraph*{Introduction.}
Cold atom experiments utilizing an optical lattice provide an excellent testbed for quantum many body 
problems which were previously inaccessible in conventional materials. A remarkable achievement is the realization 
of the Bose-Hubbard (BH) model using ultracold atoms on optical lattices~\cite{Jaksch,Greiner} with the addition 
of a confining potential that results in the ``wedding cake" structure~\cite{Batrouni2002}. Over the last two decades, a considerable 
amount of work has been devoted to understand the ground state phase diagram of the BH model and its variants
~\cite{Fisher, Batrouni1990, Krauth,Cha, Freericks, Andreas}. In general, the model contains a superfluid (SF) phase at 
incommensurate fillings and a Mott insulating (MI) phase at commensurate fillings and strong coupling.
The SF phase is gapless, whereas the MI phase is characterized by the existence of an energy 
gap for creating a particle-hole pair. As the density is changed or the interaction strength is 
varied, the BH model can be tuned from the MI to the SF. Tuning between insulating and conducting 
phases by controlling the external parameters provides a tantalizing opportunity of creating analogs to electronic devices and 
circuits by using ultra cold atoms in optical lattices, which have been recently defined as `atomtronics'~\cite{Ryu,Bhongale}. 
The conventional electronic system is based on the electron charge, whereas the atomtronic system can use neutral atoms which 
are either bosons or fermions, moreover the optical lattice is better controlled. Based on this unique property, it has been suggested 
that these atomtronic systems may be useful in quantum computing~\cite{Pepino}. Some models have already been proposed for atomtronic 
devices such as batteries, wires, diodes, and transistors~\cite{Ruschhaupt2004,Ruschhaupt2006,Ruschhaupt2008,Seaman, Pepino,Rubbo,
Stickney,Micheli,Vaishnav}.

A recent advance on optical lattices is the realization of confining potentials with toroidal shapes by using the intersection of two 
different red-detuned laser beams~\cite{Ryu, Raman}. The versatility of this technique allows the creation of ring-shaped 
lattices by superimposing an optical lattice on a toroidal confining potential, which is a realization of a quasi 
one-dimensional lattice with periodic boundary conditions. Remarkably, it is possible to control the {\it local} hopping parameter 
in a region of the ring by applying a magnetic field and an additional laser beam~\cite{Raman}. This opens up the new possibility that the 
different phases in a boson system not only can be tuned by a global parameter, such as the coupling strength or chemical potential, 
but also by a {\it local} parameter, such as the tunneling strength of a small region of the entire lattice. It has been suggested that this property can be 
utilized as an alternative realization of atomtronics~\cite{Raman}. 

In this letter, by using quantum Monte Carlo (QMC), we show that introducing weak links in a ring lattice can 
produce a local Mott (LM) phase in addition to the usual MI and SF phases present in the 
homogeneous BH model. Zero temperature local incompressible MI behavior was shown in a one dimensional system 
of interacting bosons in a confining potential~\cite{Batrouni2002}. Our non-confined model exhibits a LM 
phase which is gapless and non-SF, and a region of LM insulator which exhibits incompressible 
MI behavior. This is an important result which suggested that by controlling the local tunneling strength
the system can be tuned between a SF phase and a MI phase thorough a non-SF LM phase. This provides theoretical support 
that atomtronic switches can be implemented by tuning certain local parameters in a quasi one-dimensional system.  

\paragraph*{Model and method.}
We consider a bosonic system on a torus-shaped lattice, where the section of the torus is sufficiently 
small compared to the primary radius so the physics can be reduced to a one-dimensional lattice with periodic boundary conditions.
The Hamiltonian takes the form
\begin{equation}
\label{Hamiltonian} \nonumber \hat\mathcal H=-t \sum_{\big\langle i,j\big\rangle}w_{ij}\Big(a_i^\dagger a_j^{\phantom\dagger}+H.c.\Big)+\frac{U}{2}\sum_{i=1}^L \hat n_i(\hat n_i-1),
\end{equation}
where $L$ is the number of lattice sites.
The creation and annihilation operators $a_i^\dagger$ and $a_i^{\phantom\dagger}$ satisfy bosonic 
commutation rules, $\big[a_i^{\phantom\dagger},a_j^{\phantom\dagger}\big]=
\big[a_i^\dagger,a_j^\dagger\big]=0$, $\big[a_i^{\phantom\dagger},a_j^\dagger\big]=\delta_{ij}$,
and $\hat n_i=a_i^\dagger a_i^{\phantom\dagger}$ is the operator that measures the number of bosons on site $i$.
The parameter $t$ is the global magnitude of the hopping integral. In this paper, we use $t=1$ to set the energy scale.
The sum $\sum_{\langle i,j\rangle}$ runs over all distinct pairs of first neighboring sites $i$ and $j$, and $w_{ij}\in[0;1]$ 
determines the weakness of the hopping integral between $i$ and $j$. In the following we consider a system with $M$ 
consecutive weak links for which $w_{ij}=J/t$, where $J\in[0;t]$ is a control parameter, and $L-M$ strong links with 
$w_{ij}=1$. We restrict our study to the case with 10\% of weak links ($M=L/10$).
The parameter $U$ determines the strength of the on-site interaction.

In order to solve this model, we perform exact QMC in both canonical and grand-canonical 
ensembles by using the Stochastic Green Function  algorithm~\cite{SGF,DirectedSGF} with global space-time updates~\cite{GlobalSpaceTimeUpdate}. In the canonical ensemble,
the number of particles $N$ is a parameter and remains constant during the simulation. The chemical potential $\mu$ 
is measured at zero temperature by the finite energy difference $\mu(N)=E(N+1)-E(N)$. In the grand-canonical ensemble, 
the number of particles is given by the quantum average of the operator $\hat N=\sum_i \hat n_i$, and is controlled 
by adding to the Hamiltonian~(\ref{Hamiltonian}) the term $-\mu\hat N$ where $\mu$ is a control parameter. We use an 
inverse temperature $\beta=L/t$ in order to capture the ground-state properties.

\paragraph*{Superfluid density and compressibility.}
For the uniform system, $J=t$, only two phases are present: Mott insulator (MI) and superfluid (SF). The MI phase occurs at 
commensurate fillings and large onsite repulsion $U$, and is characterized by a vanishing compressibility, 
$\kappa=\displaystyle \frac{\partial\rho}{\partial\mu}$, where $\rho=N/L$. The SF phase is detected by measuring the 
superfluid density, $\rho_s$, given by the response of the system to a phase twist of the wave function at the boundaries. 
In our QMC simulations, 
it is convenient to relate this superfluid density to the fluctuations of the winding number, $W$, via Pollock and 
Ceperley's formula~\cite{Pollock}, $\rho_s=\displaystyle \frac{\langle W^{2}\rangle L}{2t\beta}$.
We have checked analytically and with exact diagonalization that the above formula remains valid for the non-uniform system 
where $J<t$.

\begin{figure}[h]
  \centerline{\includegraphics[width=0.5\textwidth]{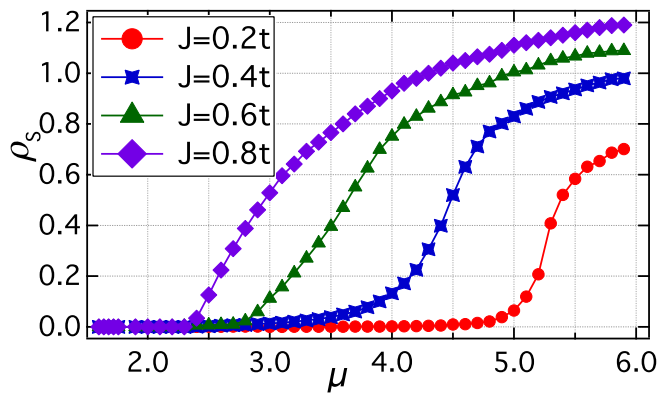}}
  \caption
    {
      (Color online) The superfluid density $\rho_s$ as a function of the chemical potential $\mu$ for $L=50$ and $U=8$ 
in the ground state.  The figure shows results for different values of the weak hopping integrals, $J=0.2t$ (circles), 
$J=0.4t$ (stars), $J=0.6t$ (triangles) and $J=0.8t$ (diamond).
    }
  \label{SF_Diff_js}
\end{figure}

In the following, we show that there exists a range of parameters for the non-uniform system for which we observe a 
vanishing superfluid density and a finite compressibility at incommensurate fillings. Fig.~\ref{SF_Diff_js} shows the 
superfluid density $\rho_s$ as a function of the chemical potential $\mu$ for $L=50$ and $U=8$. Here we use 
grand-canonical simulations for different weak link hopping $J$.  We can clearly see that the region with vanishing 
superfluid density expands over a large range of chemical potentials $\mu$ when the strength of the weak links is lowered.

\begin{figure}[h]
  \centerline{\includegraphics[width=0.5\textwidth]{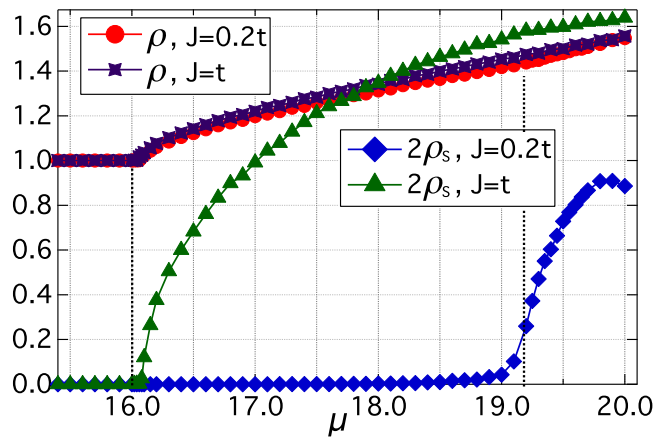}}
  \caption
    {
      (Color online) The density $\rho$ and the superfluid density $\rho_s$ as functions of the chemical potential $\mu$ for 
the homogeneous system ($J=t$) and an inhomogeneous system ($J=0.2t$), for $L=50$ and $U=20$.
    }
  \label{SF_Rho}
\end{figure}

Fig.~\ref{SF_Rho} shows the density $\rho$ and the superfluid density $\rho_s$ as functions of the chemical potential $\mu$, 
for both homogeneous ($J=t$) and inhomogeneous ($J=0.2t$) systems, for $L=50$ and $U=20$. We can see that a Mott plateau 
at $\rho=1$ exists until 
$\mu=16.1$ with a vanishing superfluid density $\rho_s$ and compressibility $\kappa$, for both systems. For $\mu>16.1$, 
the density $\rho$ starts to increase and the compressibility $\kappa$ is finite. As it is well known, the superfluid 
density $\rho_s$ of the homogeneous system is non-zero as soon as the density is no longer an integer. However, for the 
inhomogeneous system, the superfluid density remains zero until $\mu\approx 19.1$. Thus there exists a finite range of values 
for the chemical potential for which the superfluid density is vanishing but the compressibility is finite. 
Therefore, as the chemical potential is increased, the inhomogeneous system undergoes a phase
transition from a MI phase to a new phase, then to a SF phase. 

\begin{figure}[h]
\centerline{\includegraphics[width=0.5\textwidth]{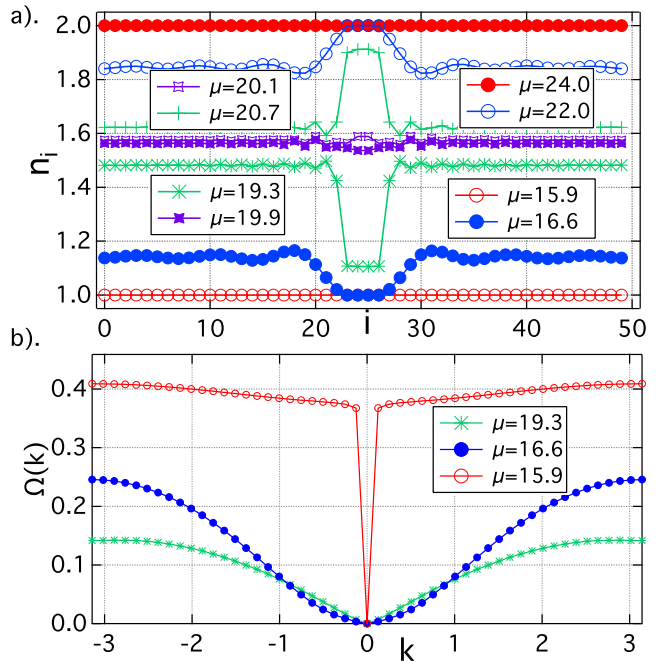}}
  \caption
    {
      (Color online) The local density (top panel) and the excitation spectrum (bottom panel) for $L=50$ and $U=20$, in the ground state.
      Top panel: The local density $n_i$ as a function of the site index $i$ for different values of the chemical potential $\mu$.
      Bottom panel: The low energy excitation spectrum $\Omega(k)$ in the three regions of the phase diagram: SF, LM and MI.
    }
  \label{Density}
\end{figure}

\paragraph*{Properties of the phases.}
We investigate the intermediate phase first by analyzing the local density of the lattice. The local density in the
homogeneous model is uniform, whether the system is in the MI or SF phase. For the inhomogeneous model, 
we have phases with a non-uniform local density, as shown in Fig.~\ref{Density}a. We insert $10$\% of the weak links in the middle 
of the lattice.  In the MI region, the local density $n_i$ throughout the entire lattice is uniform and sticks to 
integer values ($n_i=1$ for the first Mott lobe, $n_i=2$ for the second one, etc). 

When additional particles or holes are added to the lattice the weak link region keeps its integer density (see Fig.~\ref{Density}a). 
Outside the weak link region, the local density shows an oscillatory behavior. Although, based on numerical data, it is difficult to 
strictly rule out the possibility that the superfluid density is exceedingly small but non-zero in the weak link region, 
these two observations indicate that the additional particles do not affect the MI character of the 
weak link  until the number of additional particles or holes is beyond a critical density. 
For a one-dimensional system, the superfluid density or the winding is zero when part of the system is locally Mott. 
As a result we identify this locally integer-density region as a local Mott (LM) phase. The weak link 
provides an effective fixed boundary condition for the density profile, the additional particles or holes 
accumulate outside. Then, the region with $J=t$  can be 
effectively described by the hard-core boson model with $L-L_{weak}$ number of sites.  For a one dimensional 
system, the hard-core boson can be written in terms of spinless fermions using the Jordan-Wigner transformation, 
\cite{Jordan-Wigner} and the oscillation of the local density can then be explained by Friedel 
oscillations,~\cite{Friedel} where $n_{i} \sim cos(k_{F} x_{i})$, where $k_{F}$ is the Fermi wavevector given 
by the particle density. This explanation is corroborated by the numerical data which show that 
the cycle of the oscillation of the local density is approximately given by $1/|n-1.0|$ for $\mu=16.6$ and $1/|n-2.0|$ 
for $\mu=22$ (see Fig.~\ref{Density}a).

When adding more particles beyond the critical density, the local density at the weak link shifts 
away from integer values. This suggests that the LM insulating region is destroyed. Thus it opens the path 
for the flow, and we find that the superfluid density becomes finite when this happens. 

We study the dynamics of the model by evaluating the low energy excitation spectrum.
Using the Feynman single-mode approximation, the low energy excitation spectrum $\Omega(k)$ can be written as~\cite{Scalettar}
\begin{equation}
 \label{EnergySpectrum} \Omega(k)=\frac{E_{k}}{S(k)}
\end{equation}
where, 
\begin{equation}
 \label{EnergyS}  E_{k}=\frac{-t}{L}(\cos k-1)\langle \Psi_{0}|\sum_{ i=1}^{L}(a_i^\dagger a_{i+1} +a_{i+1}^\dagger a_i)|\Psi_{0}\rangle,
\end{equation}
$|\Psi_{0}\rangle$ is the ground state, and $S(k)$ is the static structure factor.

Fig.~\ref{Density}b displays the low energy excitation spectrum throughout the reciprocal 
lattice space. In the MI region it shows a gap near zero wave vector, whereas it has a linear dependence for the SF phase. The linear 
behavior is expected in the SF region due to the gapless Goldstone mode. In the LM region the low energy spectrum shows a 
parabolic behavior, as expected for disordered free-particles. Since the LM does not follow a linear behavior near $k=0$, no signal 
of super-flow  exists in the LM region. 

\begin{figure}[t]
  \centerline{\includegraphics[width=0.5\textwidth]{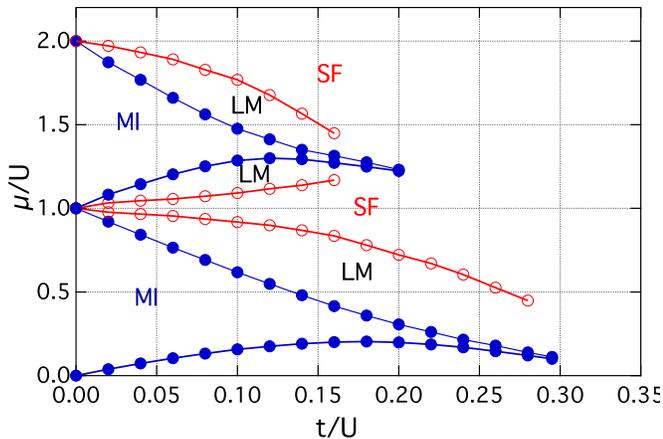}}
  \caption
    {
      (Color online) The ground state phase diagram of the inhomogeneous system ($J=0.2t$) in the $(\mu/U,t/U)$ plane. The lines 
with solid circles show the first and the second Mott lobes, and the lines with open circles show the boundaries between the 
LM and the SF regions.
    }  
  \label{Phase}
\end{figure}

\paragraph*{Ground state phase diagram.}
The MI phase is characterized by an integer local density and the existence of a finite gap for single particle excitations.
At zero temperature, the gap can be easily obtained in the canonical ensemble. We define the gap as $\Delta=\mu_+-\mu_-$, where $\mu_-$ and 
$\mu_+$ are the minimum and maximum values of the chemical potential for which the MI phase exists. By definition, $\mu_+=E(N+1)-E(N)$ 
and $\mu_-=E(N)-E(N-1)$, where $N$ is the number of particles in the MI phase. The functions $\mu_-(t,J,U)$ and $\mu_+(t,J,U)$ determine 
the boundaries between the MI and LM regions. Since the total density remains unchanged for $\mu\in[\mu_-;\mu_+]$, the compressibility
$\kappa$ is vanishing in the MI region.

We determine the phase boundary between the LM and SF regions by using the grand canonical simulations and scanning over the chemical 
potential, as in Fig.~\ref{SF_Rho}. The critical value $\mu_c$ of the chemical potential where the superfluid density becomes non-zero 
depends on the size of the system, $L$, and converges to a finite value in the thermodynamic limit. As the size increases, the curve 
displaying the superfluid density becomes sharper and sharper. Since we work with a fixed large size, $L=50$, we define $\mu_c$ by the 
value of the chemical potential that corresponds to the maximum slope for the superfluid density curve. The curve $\mu_c(t,J,U)$ 
determines the boundary between the LM and SF phases.

In our simulations the density varies continuously as a function of the chemical potential. This suggests that the transitions from 
MI to LM and from LM to SF are continuous, as it is the case for the homogeneous model~\cite{Till,Elesin,Kash}.
We show in Fig.~\ref{Phase} the ground state phase diagram for $J=0.2t$ in the $(\mu/U,t/U)$ plane. The Mott lobes that are present 
in the homogeneous model are weakly deformed by the presence of the LM phase. The phase boundaries near the tip of the Mott lobes 
are difficult to estimate due to the very small LM region.

\begin{figure}[t]
  \centerline{\includegraphics[width=0.5\textwidth]{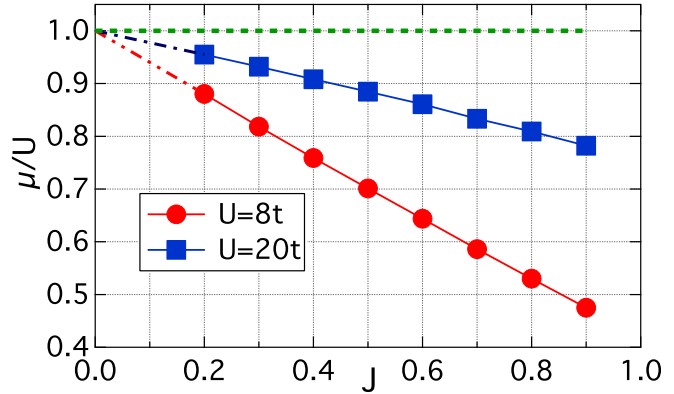}}
  \caption
    {
      (Color online) The critical value of the chemical potential $\mu_c$ between the LM and SF regions, as a function of the weak 
link hoping $J$, for $U=8$ and $U=20$. The variation with $J$ is quasi-linear.
    } 
  \label{Mu_j}
\end{figure}

We now investigate the variation of the phase boundary between the LM and SF regions as a function of weak link hopping $J$, 
Fig.~\ref{Mu_j}. For a fixed value of the interaction $U$, the phase boundary lifts up linearly when decreasing the hopping $J$ 
in the weak link reducing the size of the SF region in the phase diagram. In the limit $J=0$ the curve extrapolates to $\mu/U=1$, 
and the SF region completely disappears since the system is no longer periodic. 
\paragraph*{Conclusion.}
In this study, we propose that a superconducting ring with weak links might display a phase which is gapless, compressible, 
and non-superfluid, with local Mott insulating behavior. This phase does not exist in the homogeneous Bose-Hubbard model. 
We expect that in the thermodynamic limit, the weak link acts effectively as a domain wall which suppresses the superfluid. 
While a thorough characterization of the phases and the critical properties of the model will require an analysis of the 
inhomogeneous Luttinger liquid coupled to a lattice, which is an interesting challenging topic by itself\cite{Rech,Safi,Maslov,Ponomarenko}, 
we hope our work motivate further study in this direction. 
Perhaps the most important aspect of the present study is to understand the mechanism of controlling superfluid flow by 
local perturbation on a finite size system, which is directly related to atomtronic.
In the experiment by Raman et al., a toroidal condensate is created with a smooth trapping potential~\cite{Raman}. 
If the experiment can be repeated by superimposing a lattice on top of the toroidal potential, our model could be directly studied experimentally. 
Our results have direct implications for atomtronic devices~\cite{Seaman, Pepino}.  
For example, if the chemical potential $\mu$, the weak link hopping $J$, and the interaction $U/t$ are tuned 
so that only the link is a Mott insulator, then a gate above the link can be used to switch 
the conductivity of the link on and off. The non-linearity of the switching can be tuned by adjusting the link width and hoping J/t. 
Complex circuits with highly non-linear behavior may be constructed by a series of such switches.  

\paragraph*{Acknowledgments.}
We would like to acknowledge Daniel Sheehy for suggesting this problem to us and we thank him and  K. C. Wright for
useful discussions. This work is supported by NSF OISE-0952300 (KH, VGR and JM) and by DOE SciDAC grant DE-FC02-06ER25792 (KMT and MJ). This work used 
the Extreme Science and Engineering Discovery Environment (XSEDE), which is supported by the National Science Foundation grant number 
DMR100007, and the high performance computational resources provided by the Louisiana Optical Network Initiative (http://www.loni.org).

\end{document}